\documentclass{article}

\usepackage[utf8]{inputenc} 
\usepackage[T1]{fontenc}    
\usepackage{hyperref}       
\usepackage{url}            
\usepackage{booktabs}       
\usepackage{amsfonts}       
\usepackage{nicefrac}       
\usepackage{microtype}      
\usepackage{xcolor}         
\usepackage{listings} 
\usepackage{amssymb}
\usepackage{a4wide}

\newcommand{\R}{{\mathbb R}}
\newcommand{\X}{{\mathbf x}}
\newcommand{\Y}{{\mathbf y}}

\title{Differentiable Scripting}

\author{%
  Uwe~Naumann \\
  Department of Computer Science, RWTH Aachen University\\
  52056 Aachen, Germany \\
  \texttt{naumann@stce.rwth-aachen.de}
}

\date{}

\begin{document}

\maketitle

\begin{abstract}
In Computational Science, Engineering and Finance (CSEF)
scripts typically serve as the ``glue'' between potentially highly 
complex and computationally expensive external subprograms.
Differentiability of the resulting programs turns out to be
essential in the context of derivative-based methods for error analysis, uncertainty quantification, optimization or training of surrogates.
We argue that it should 
be enforced by the scripting language
itself through exclusive support of differentiable (smoothed) external 
subprograms and differentiable intrinsics combined with
prohibition of nondifferentiable branches in the data flow.
Illustration is provided by a prototype adjoint code compiler for a
simple Python-like scripting language.
\end{abstract}

\section{Motivation}

In an ideal world, libraries for numerical simulation would provide a 
complete collection of the state of the art in the target domain 
implemented as easy-to-use, efficient, scalable and sustainable software ... 
and corresponding adjoints. Considerable
progress in this direction has been made over the past three decades thanks 
to activities in Algorithmic (also: Automatic) Differentiation (AD) 
\cite{Griewank2008EDP} including software tool development
and applications in CSEF
as well as evolution of its theoretical foundations.
A good overview is provided by the proceedings of so far seven
international AD conferences, e.g. \cite{Bischof2008AiA,Christianson2018Sio,Forth2012RAi}. 
See \url{www.autodiff.org} for links to 
research groups and software tools in addition to an extensive bibliography 
on the subject.

Numerical simulation in CSEF relies on a hierarchy of 
software library components implementing
multivariate vector functions
$
F : \R^n \rightarrow \R^m : \Y=F(\X)
$ (also: primals)
including subprograms for linear algebra, model calibration, 
Monte Carlo simulation and solvers for algebraic and differential equations. 
Substantial effort has been put into the implementation of 
adjoints
$
\bar{F} : \R^n \times \R^n \times \R^m \rightarrow \R^n : \bar{\X}=\bar{F}(\X,\bar{\X},\bar{\Y}) \equiv \bar{\X} + F'(\X)^T \cdot \bar{\Y} 
$
while ensuring or at least assuming differentiability of $F$ at all points of 
interest.
Very large implementations in C/C++, Fortran and
other general-purpose programming languages can be handled by state of the art 
AD software tools. Both source-to-source transformation \cite{Giering1998RfA} 
and operator/function
overloading combined with template metaprogramming techniques 
\cite{Phipps2012EET} have been 
developed and applied successfully for several decades.
Cutting-edge solutions run on large CPU, e.g. \cite{Heimbach2005Aee}, or GPGPU 
clusters, e.g. \cite{Gremse2015GAA}. 

Simulation scenarios of interest are often synthesized from a set of
differentiable subprograms using both syntactically and semantically 
relatively simple instructions provided by domain-specific scripting languages. 
The latter typically do not require the 
same complexity exhibited by general-purpose 
programming languages.
Their users should not be required to have the same level of software 
development expertise as their colleagues who implement the core 
numerical libraries. Simple flow of control 
over the basic arithmetic and relational operators in addition to
a set of differentiable intrinsics / subprograms often 
suffices. The automatic generation of adjoint scripts becomes relatively 
straight forward in this case. 

Conditions for differentiable scripting are formulated in the following section.
Their integration into a single-pass adjoint code compiler for a simple 
Python-like scripting language is the subject of Section 3. A case study 
is presented in Section 4 followed by final remarks in Section 5.

\section{Differentiable Scripting}

A scripting language SL is differentiable if
all arithmetic operations and intrinsic functions as well as all callable 
external subprograms are differentiable and if
active branches in the flow of control are prohibited.
For illustration we introduce a simple imperative Python-like\footnote{We use a tiny subset of Python and curly brackets instead of ``pythonic'' indentation.} scripting language 
supporting collections of (internal) subprograms
\begin{lstlisting}
subprogram : DEF NAME '(' arguments ')' '{' statements RETURN argument '}'
\end{lstlisting}
over statements defined as
\begin{lstlisting}
statement : assignment | if | while | for | call | ...
\end{lstlisting}
with assignments 
\begin{lstlisting}
assignment: variable '=' expression
\end{lstlisting}
of the results of right-hand sides of type
\begin{lstlisting}
expression: expression '*' expression | EXP '(' expression ')' | ...
\end{lstlisting}
to scalar variables. Intraprocedural flow of control includes
conditional branches
\begin{lstlisting}
if : IF condition '{' statements '}' ELSE '{' statements '}' 
\end{lstlisting}
and two types of loops.
Internal as well as external subprograms can be called as
\begin{lstlisting}
call: NAME '=' NAME '(' arguments ')'
\end{lstlisting}
Active in- and outputs of subprograms need to be vectors. This restriction
(as well as others) is likely to be lifted in upcoming versions of SL. Its
development should be considered as work in progress.

We present a single-pass compiler which translates SL into Python. 
It is implemented in C-style C++ 
using the scanner and parser 
generators {\tt flex} and {\tt bison} \cite{flexbison}. 
Grammar rules are specified in {\tt bison} syntax. [Non-]Terminal symbols are
written in [lower-]uppercase letters. The entire source code can be found on
\url{github.com/un110076/AD4SL}including a complete specification of the 
SL grammar. All results in this paper can thus be reproduced.

\subsection{Differentiable Intrinsics}

The differentiable arithmetic operations \lstinline{+,-,*,/} over scalar 
floating-point variables are supported as well as commonly used differentiable
arithmetic functions such as the exponential function represented by the
terminal symbol \lstinline{EXP}. As an example for a smoothed 
not everywhere differentiable intrinsic we consider $max(x,0)$ implemented 
as part of the SL runtime library in C++\footnote{pybind11 ({\tt github.com/pybind/pybind11}) is used to link the SL runtime library with the adjoint 
script.} as
\begin{lstlisting}
float gt0(float x) { return fmax(x,0.0); }
\end{lstlisting}
Differentiability at $x=0$ can be ensured, for example, by sigmoidal
smoothing yielding the following implementation of the derivative of \lstinline{gt0} with respect to its input argument:
\begin{lstlisting}
float d_gt0(float x) {
  const float h=1e-3;
  if (x<-h) return 0; else if (x>h) return 1;
  else return 1./(1.+exp(-(x)/h));
}
\end{lstlisting}
The change in tangent slope at $x=0$ is globalized in dependence of
the hyperparameter \lstinline{h}. Suitable choices for 
values of \lstinline{h} depend on the simulation to be implemented.
Less static ways of specifying the value of \lstinline{h} may be required
in practice.
Discontinuities can be smoothed similarly. Alternatively, central finite
difference approximations of the derivative can be implemented. Moreover, there
is a rich body of published work to be considered, e.g. \cite{Griewank2013Osp,Kearfott1996IEo,Khan2013Eae}. For the purpose of this paper, all external 
subprograms and intrinsics are assumed to be (made) differentiable independent
of the actual smoothing technique.
Nondifferentiable intrinsics might still be permitted.
If one is evaluated at (or within a 
neighborhood of) a point of nondifferentiability, then an exception
should be raised and handled appropriately. At the very least, users
should be informed about nondifferentiability in order to
not loose sight of potentially serious numerical implications.

\subsection{Passive Branches}

To guarantee differentiability of an SL script,
branch conditions must not depend on active variables. 
This constraint can be enforced by static activity analysis as described in
\cite{Hascoet2005TBR}. The iterative nature of static data flow analysis 
makes it unsuitable for syntax-directed (i.e., single-pass)
compilation. Instead, variables in SL are partitioned into active and 
passive using 
names starting with lowercase and uppercase letters, respectively. This 
conservative approach allows for propagation of the potential activity of 
expressions
by augmenting the SL grammar with a synthesized attribute 
for forward activity, referred to as {\em variedness} in \cite{Hascoet2005TBR}. 
Attributes of terminal symbols (leafs of the parse tree) are initialized 
according to the case of their respective first letters followed by 
conjunctive bottom-up propagation, for example,
\begin{lstlisting}
expression : expression '*' expression { $$.v=$1.v||$3.v; }
\end{lstlisting}
where nonterminal symbols are referenced by their position within the
right-hand side of the production rule 
(\lstinline{$1} and \lstinline{$3}) and 
the left-hand side is accessed by \lstinline{$$}. For example, variedness of
the first term is indicated by \lstinline{$1.v==true}.

The actual treatment of active branches as errors or mere warnings is up to
the language designers. In our prototype we use assertions (from \lstinline{<cassert>}),
for example,
\begin{lstlisting}
if : IF condition '{' statements '}' { assert(!$2.v); } ELSE ...
\end{lstlisting}

\section{Single-Pass Adjoints}

Adjoint SL code generated by our single-pass compiler follows the state 
of the art in adjoint code generation according to the well-known 
AD principles implemented, for example, by Tapenade \cite{Hascoet2013TTA}. 
To implement the fundamental requirement of data flow reversal,
adjoint SL subprograms consist of forward (sythesized in the string attribute 
\lstinline{$$.f}) and reverse (in \lstinline{$$.s} and \lstinline{$$.a}) 
sections.
The forward section corresponds to the primal code (in \lstinline{$$.p}) 
including
storage of required values that are overwritten. Some primal statements
may not be required for the evaluation of the adjoint. Their generation as
part of the forward section can be suppressed explicitely by using 
\lstinline{#pragma noprimal}. Similarly, storage as well as subsequent recovery
of overwritten values which are
not required by the adjoint can be avoided by using \lstinline{#pragma notbr}.
The adjoint code remains correct if pragmas are not used. 
Memory requirement and runtime efficiency may suffer though. Incorrect 
placement of pragmas results in erroneous adjoint code.

In the reverse section all assignments are decomposed into single-assignment 
code (sac; in \lstinline{$$.s}) through storage of the results of all 
operations
in locally unique program variables. Corresponding adjoint statements are
synthesized in \lstinline{$$.a}. See Section 4 for an example.

Syntax-directed translation \cite{dragoonBook} is a conceptually elegant and light-weight approach
to source-to-source transformation for typically relatively simple 
scripting languages. Their grammar can be augmented with 
attributes for synthesis of adjoint code during a single pass of a 
shift-reduce parser. Explicit construction of an intermediate representation 
of the primal program ist not required. A compact, easily
maintainable source-to-source transformation engine can be built with low effort
at the expense of 
not being able to perform iterative static program analysis. This compromise
turns out to be reasonable under the assumption that the potential for
code optimization at the scripting level is low.

Adopting the approach proposed in \cite{Naumann2012TAo}
we use {\tt flex}/{\tt bison} to generate a generalized
operator-precedence LR(1) parser for a (not quite; see below) S-attributed 
grammar for SL. Production rules are augmented with semantic actions
over those attributes, for example
\begin{lstlisting}[numbers=right]
assignment : variable '=' { j=0; } expression {
    assert($1.v||!$4.v);
    $$.p=$1.p+'='+$4.p;
    if (!notbr&&!noprimal) {
      $$.f="push_s("+$1.p+')';
      $$.a=$1.p+"=pop_s()";
    } else { $$.f=""; $$.a=""; notbr=false; }
    if (!noprimal) $$.f+=$1.p+'='+$4.p; else noprimal=false;
    if ($$.v)
      $$.a+=$4.s A($4.j)+'='+A($4.j)+'+'+adj+$1.p adj+$1.p+"=0.0"+$4.a;
  }
\end{lstlisting}
An inherited attribute (\lstinline{$4.j}) is used to enumerate the sac 
variables in the right-hand sides of assignments. It is implemented as a global counter \lstinline{j} which is
reset to zero prior to parsing a new right-hand side (see line 1). Assignment of active values to passive variables is not permitted (line 2). The primal
code is unparsed as is (line 3). It is also used in the forward section of 
the adjoint code preceded by storage of the value of the left-hand side (lines 5 and 8). Optionally, both the primal assignment and/or storage and recovery
of its left-hand side can be omitted through use of the previously 
introduced pragmas. 
Again, the corresponding static dead code \cite{dragoonBook} and 
to-be-recorded \cite{Hascoet2005TBR} analyses do not comply with the 
single-pass compilation paradigm. 
Adjoint code is generated for active assignments 
exclusively (lines 9 and 10).
The stack handlers (\lstinline{push_s}, \lstinline{pop_s} in addition to
vector versions and stack access for reversal of flow of control) are part of 
the SL runtime library.

Right-hand sides of assignments are
defined recursively as expressions resulting from arithmetic operations or 
intrinsic functions applied to (sub-)expressions. 
Their sac is synthesized in \lstinline{$$.s}.
Corresponding adjoint statements are generated in \lstinline{$$.a}.
For example, scalar multiplication is implemented as follows:
\begin{lstlisting}[numbers=right]
expression : expression '*' expression {
    $$.j=j++; 
    $$.p=$1.p+'*'+$3.p;
    $$.s=$1.s+$3.s+V($$.j)+'='+V($1.j)+'*'+V($3.j);
    $$.a="";
    if ($1.v) $$.a+=A($1.j)+'='+A($1.j)+'+'+V($3.j)+'*'+A($$.j);
    if ($3.v) $$.a+=A($3.j)+'='+A($3.j)+'+'+V($1.j)+'*'+A($$.j);
    $$.a+=A($$.j)+"=0.0";
    if ($$.v) $$.a+=$3.a+$1.a;
  }
\end{lstlisting}
Each reduction of a handle (right-hand side of a production rule) triggers
the evaluation of the three synthesized string attributes 
\lstinline{$$.p}, \lstinline{$$.s} and \lstinline{$$.a}.
Enumeration of sac variables requires the
inherited attribute \lstinline{$$.j} (line 2). 
Two auxiliary functions generate sac variables (\lstinline{V}) 
and their adjoints (\lstinline{A}). 
Adjoints of active sac variables are incremented (lines 6 and 7). Hence,
they must be reset to zero after use as a left-hand side of an assignment
(line 8). 
Refer to lines 6--10 of the adjoint code in Section~\ref{sec:cs} 
for an example.

The flow of control is reversed by flagging branches and by counting loop
iterations as suggested in \cite{Naumann2012TAo}.
Within an \lstinline{if} statement, execution of the 
\lstinline{if} branch yields storage of $1$ (true). The 
\lstinline{else} branch is marked by $0$ (false). 
Loop iterations are counted by incrementing counter. The
adjoint loop body is executed as often as the primal loop body.
Simple \lstinline{for} loops can be reversed explicitly; see Section 4. Nesting is supported.
Subprogram calls may require storage (\lstinline{push_v}) and recovery 
(\lstinline{pop_v}) of values overwritten by their results. 
\begin{lstlisting}
call : NAME '=' NAME '(' arguments ')' {
  $$.p=$1.p+'='+$3.p+'('+$5.p+") ";
  if (!notbr&&!noprimal) {
    $$.f="push_v("+$1.p+')';
    $$.a=$1.p+"=pop_v()";
  } else { $$.f=""; $$.a=""; notbr=false; }
  if (!noprimal) $$.f+=$1.p+'='+$3.p+'('+$5.p+") "; else noprimal=false;
  $$.a+=$5.s+'='+adj+$3.p+'('+$5.a+')';
}
\end{lstlisting}
Active arguments are augmented with corresponding adjoint arguments.
The effect of using the aforementioned pragmas is similar to the 
\lstinline{assignment} case.
Indentation (corresponding code omitted from listings) is handled correctly 
by the compiler.

\section{Case Study} \label{sec:cs}

As a case study we consider a Monte Carlo pricer for a European call 
option \cite{Hull2018OFa} implemented in SL as follows:
\begin{lstlisting}[language=Python,numbers=right]
from numpy import *
from intrinsic import *
from external import *

def payoff (d,p) {
  #pragma noprimal
  p[0]=gt0(d[1]-d[0])
  return p
}

def black_scholes_call (x,y,M) {
  s=[0.0]*M
  #pragma notbr
  s=mc(x,s,M)
  d=[0.0]*2
  p=[0.0]*1
  for I in range(M) {
    #pragma notbr
    d[0]=x[3]
    d[1]=s[I]
    #pragma notbr
    p=payoff(d,p)
    #pragma notbr
    y[0]=y[0]+exp(-x[1])*p[0]
  }
  #pragma noprimal
  y[0]=y[0]/M
  return y
}
\end{lstlisting}
The function \lstinline{black_scholes_call} takes four active inputs collected
in the array \lstinline{x} consisting of asset price, interest rate, volatility and strike. \lstinline{M} sample paths are allocated in line 12 and simulated 
by calling the function 
\lstinline{mc} in line 14. The Monte Carlo simulation 
of the Black-Scholes stochastic differential equation \cite{black1973pricing} 
is implemented in C++. 
The corresponding adjoint Monte Carlo simulation 
can be generated by your favorite AD tool. 
We provided a hand-written adjoint.
Both are part of \lstinline{external} (line 3).
Vectors \lstinline{d} and \lstinline{p} need to be allocated (lines 15 and 16) prior to
calling \lstinline{payoff} (line 22).
Discounted local payoffs are added up for all 
paths in line 24 and averaged to get an estimate for the overall payoff 
\lstinline{y[0]} in line 27.
The function \lstinline{payoff} reduces to calling the 
previously discussed intrinsic \lstinline{gt0} in line 7 with smoothed 
derivative provided by \lstinline{intrinsic} (line 2). The primal code does not 
contain any active branches.

The \lstinline{noprimal} pragma is applied to the last assignments in each 
subprogram, respectively (lines 6 and 26). Overwrites of \lstinline{s} (line 14), \lstinline{d[0]} (line 19), \lstinline{p} (line 22), and \lstinline{y[0]} (in lines 24 and 27) do not result in 
loss of required values making \lstinline{#pragma notbr} applicable. 

The compiler transforms the primal code into the following adjoint:
\begin{lstlisting}[language=Python,numbers=right]
def a_payoff(d,a_d,p,a_p) :
  v=[0.0]*4; a_v=[0.0]*4
# forward section
    # p[0]=gt0(d[1]-d[0]) omitted due to #pragma noprimal
# reverse section
    # sac
    v[0]=d[1]; v[1]=d[0]; v[2]=v[0]-v[1]; v[3]=gt0(v[2]) 
    # adjoint sac
    a_v[3]=a_v[3]+a_p[0]; a_p[0]=0.0
    a_v[2]=a_v[2]+d_gt0(v[2])*a_v[3]; a_v[3]=0.0
  ...
  return a_d

def a_black_scholes_call(x,a_x,y,a_y,M) :
  ...
# forward section
  s=[0.0]*M; a_s=[0.0]*M
  # push_v(s) omitted due to #pragma notbr
  s=mc(x,s,M)
  ...
  for I in range(M) :
    ...
    push_s(d[1])
    d[1]=s[I]
    # push_v(p) omitted due to #pragma notbr
    p=payoff(d,p)
    # push_v(y) omitted due to #pragma notbr
    y[0]=y[0]+exp(-x[1])*p[0]
  # y[0]=y[0]/M omitted due to #pragma noprimal
# reverse section
  v[0]=y[0]
  ...
  a_y[0]=a_y[0]+a_v[0]
  a_v[0]=0.0
  for I in reversed(range(M) ) :
    ...
    d[1]=pop_s()
    # p=pop_v() omitted due to #pragma notbr
    a_d=a_payoff(d,a_d,p,a_p)
    ...
  # s=pop_v() omitted due to #pragma notbr
  a_x=a_mc(x,a_x,s,a_s,M)
  return a_x
\end{lstlisting}
Adjoint versions of both subprograms are generated. 
The listing was edited to comply with the overall space restrictions and to add
crucial comments. In 
\lstinline{a_payoff} the primal intrinsic \lstinline{gt0} is omitted
in the forward section (line 4). Its smoothed adjoint is called
in the reverse section (line 10). Four sac variables and their adjoints 
are required (line 2).

The results of the \lstinline{M} primal Monte Carlo path simulations
are augmented with storage for the corresponding adjoints in line 17.
Reversal of the simple for loop in line 21 does not require additional 
information to be stored in the forward section. 

The reverse section starts in line 30 with the adjoint of line 27 of 
the primal code.
The adjoint simple for loop in line 35 iterates backwards over the 
Monte Carlo paths thus ensuring correct recovery of all required values 
(of \lstinline{d[1]} stored in line 23). 
Finally, the adjoint Monte Carlo simulation is called in line 42
yielding the four adjoints of interest in \lstinline{a_x}.
Several intermediate statements are omitted due to space restriction. 

\section{Final Remarks}

Differentiability of scripts for CSEF can and should be 
enforced by the scripting language
to prevent non-expert users 
from unintentional implementation of nondifferentiable programs.
The problem is shifted to a large extent to the
design of a library of differentiable intrinsics and differentiable
callable external subprograms. At the very least, ``differentiation-aware programming'' should be facilitated.

\end{document}